\documentclass[12pt]{article}
\newcommand{\be}{\begin{equation}}
\newcommand{\ee}{\end{equation}}
\newcommand{\beas}{\begin{eqnarray*}}
\newcommand{\eeas}{\end{eqnarray*}}
\newcommand{\bea}{\begin{eqnarray}}
\newcommand{\eea}{\end{eqnarray}}
\newcommand{\ba}{\begin{array}}
\newcommand{\ea}{\end{array}}
\newcommand{\nn}{\nonumber}

\newcommand{\al}{\alpha}

\newcommand{\si}{\sigma} 

\newcommand{\mP}{{\rm P}}

\begin{document}
\title{
\vspace*{-4ex}
{\bf 
Accelerated expansion of the Universe\\
filled up with the scalar gravitons
}}
\author{Yu.\ F.\ Pirogov
\\
\it Theory Division, 
Institute for High Energy Physics,  Protvino, \\
\it RU-142281 Moscow Region, Russia
}
\date{}
\maketitle
\abstract{\noindent 
The concept of the scalar graviton as the source of the dark matter
and dark energy of the gravitaional origin is applied to study the
evolution of the isotropic homogeneous Universe. 
A~realistic self-consistent solution to the modified pure gravity
equations, which correctly describes
the accelerated expansion of the spatially flat Universe, is found and
investigated. It is argued that the scenario with the scalar
gravitons filling up the Universe may emulate the LCDM model, reducing
thus the true dark matter to an artefact.}


\section{Introduction}

According to the present-day cosmological paradigm our Universe is
fairly isotropic, homogeneous, spatially flat and experiences
presently the accelerated expansion. The conventional description of
the latter phenomenon is given by the model with the
$\Lambda$-term and the cold dark matter (CDM).\footnote {Hereof, the
LCDM model. For a review on cosmology, see, e.g.,
ref.~\cite{astro}.} Nevertheless, such a
description may be just a phenomenological reflection of a more
fundamental mechanism. A realistic candidate on such a role is
presented in the given paper.

In a preceding paper~\cite{Pir1}, we proposed a modification of the
General Relativity (GR), with the massive scalar graviton in addition
to the massless tensor one.\footnote{For a brief exposition of such a
modified GR, see ref.~\cite{Pir2}.}  The scalar graviton was put
forward as a source of the dark matter (DM) and the dark energy (DE)
of the gravitational origin. In ref.~\cite{Pir3},
this concept was applied to
study the evolution of the isotropic homogeneous Universe. The
evolution equations were derived and the plausible arguments in favour
of the reality of the evolution scenario with the scalar gravitons
were presented. 

In the present paper, we expose an explicit solution to the evolution
equations in the vacuum, which gives
the correct description of the accelerated expansion of the spatially
flat Universe.  It is shown that
the emulation of the LCDM model can indeed be reached
as it was anticipated earlier~\cite{Pir3}. In Section~2, we first
briefly remind the evolution equations in the vacuum filled up only
with the scalar gravitons. Then the master equation for the Hubble
parameter is presented. Finally, a self-consistent solution of the
latter equation, possessing  the desired properties, 
is found and investigated. In the Conclusion, the proposed
solution to the DM and DE problems is recapitulated.

\section{Accelerated expansion}

\paragraph{Evolution equations}

We consider the isotropic homogeneous Universe without the true DM.
Besides, we neglect by the luminous matter missing thus the initial
period of the Universe evolution. Then, the
vacuum evolution equations look like\footnote{We refer the reader to
ref.~\cite{Pir3} for more details.}
\bea\label{eveq}
3\bigg(\Big(\frac{\dot a}{a}\Big)^2+ \frac{\kappa^2}{a^2}\bigg)
&=&\frac{1}{m_{\rm P}^2} (\rho_{\rm s}+ 
\rho_{\rm \Lambda})   ,\nn\\
2\, \frac{\ddot a}{a}
+\Big(\frac{\dot a}{a}\Big)^2 
+ \frac{\kappa^2}{a^2} 
&=&-\frac{1}{m_{\rm P}^2}  (p_{\rm s}+ 
p_{\rm \Lambda})  ,
\eea
with $a(t)$ being the dynamical scale
factor of the Universe, $t$ being the comoving time and $\dot
a =d a/d t$, etc. In the above, $\kappa^2$ is proportional to the
spatial curvature, with $\kappa^2=0$ for the spatially flat Universe. 
The parameter $m_{\rm P}$ is the Planck mass.

On the r.h.s.\ of eq.~(\ref{eveq}), $\rho_\Lambda$ and  $p_\Lambda$
are the energy density and the pressure 
corresponding to the cosmological constant $\Lambda$:
$\rho_\Lambda=-\rho_\Lambda=m_{\mP}^2\Lambda\geq 0$.
Likewise, $\rho_{\rm s}$ and  $p_{\rm s}$ are, respectively, the
energy density and pressure of the scalar gravitons:  
\bea\label{rhop}
\rho_{\rm s} &=&  
f_{\rm s}^2\Big( \frac{1}{2}\dot\si^2  +
3 \frac{\dot a}{a}\dot\si+\ddot\si\Big )+ m^2_\mP
\Lambda_{\rm s}(\si),\nn\\
p_{\rm s}  &=&  
f_{\rm s}^2\Big(\frac{1}{2}\dot\si^2  -
3 \frac{\dot a}{a}\dot\si-\ddot\si \Big)-m^2_\mP \Lambda_{\rm s}(\si). 
\eea
Here, $f_{\rm s}={\cal O}(m_{\rm P})$ is a constant with the
dimension of mass entering the kinetic term of the scalar graviton
field $\si$. The latter in the given context looks like
\be
\si=3\ln \frac{a}{\tilde a},
\ee
with ${\tilde a(t)}$ being a nondynamical scale factor given a priori.
The $\si$-field is defined up to an additive constant.
Without any loose of generality, we can fix the constant
by the asymptotic condition: $\si(t)\to 0$ at $t\to \infty$. 

In eq.~(\ref{rhop}), we put
\be
V_{\rm s}+\partial V_{\rm s}/\partial \si\equiv
m^2_\mP\Big(\Lambda_{\rm s}(\si)+\Lambda\Big),
\ee
where $V_{\rm s}(\si)$ is the scalar graviton potential. 
More particularly, we put
\be
V_{\rm s}=V_0+\frac{1}{2}m_{\rm s}^2
f_{\rm s}^2(\si-\si_0)^2+{\cal O}\Big((\si-\si_0)^3 \Big),
\ee
with $\si_0$ being a constant,
$f_{\rm s}(\si-\si_0)$ the physical field of the scalar graviton
and $m_{\rm s}$ the mass of the latter. By their  nature,
$\Lambda_{\rm s}$ and $\Lambda$ are quite similar.
To make the division onto
these two parts unambiguous we normalize 
$\Lambda_{\rm s}$ by an additive constant so that 
$\Lambda_{\rm s}(0)=0$.   Clearly, we get from
eq.~(\ref{rhop}) that
$\rho_{\rm s}+p_{\rm s}=f_{\rm s}^2 \dot\si^2$. Here, 
the contribution of $\Lambda_{\rm s}$ exactly cancels what is 
quite similar to the relation $\rho_\Lambda+p_\Lambda=0$. So, the
contribution of $\Lambda_{\rm s}$ is a kind of the dark energy.
In what follows,
we put $\si_0=0$ and $V_0=m_{\rm P}^2\Lambda$, with  $\si\to 0$ at
$t\to\infty$ becoming the ground state.

The nondynamical functions $V_{\rm s}$ and $\tilde a$ being the two
characteristics of the vacuum are not quite independent.
More particularly, adopting the isotropic homogeneous
ansatz for the solution of the
gravity equations, with only one dynamical variable~$a$, we tacitly
put a consistency relation between  $\tilde a$ and
$V_{\rm s}$. As a result,  only 
one combination of the two  lines of eq.~(\ref{eveq}) is the  true
equation of evolution, with the second independent combination giving
just the required consistency condition.

\paragraph{Master equation}

In what follows, we restrict ourselves by the case of the spatially
flat Universe, $\kappa=0$. Subtracting the  first line of
eq.~(\ref{eveq}) from the
second one and accounting for eq.~(\ref{rhop}) we get the relation
\be\label{Hsi}
\dot H=-\frac{1}{4}\al \dot \si^2,
\ee
where $ H\equiv \dot a/a$ is the Hubble parameter and 
\be
\al=2\bigg(\frac{f_{\rm s}}{m_\mP}\bigg)^2.
\ee
We assume that $\al ={\cal O}(1)$. Substituting $\dot \si$ given by
eq.~(\ref{Hsi}) into the first line of eq.~(\ref{eveq}) we get the
integro-differential master equation for the Hubble parameter:
\be\label{meq}
H^2= -\frac{1}{3}\Big(\frac{\sqrt{\al}}{2}
\frac{\ddot H +6H \dot H}{\sqrt{-\dot H}} +
\dot H\Big)
+\frac{1}{3}\Big(\Lambda_{\rm s}(\si)+\Lambda\Big).
\ee
where it is to be understood
\be\label{intsi}
\si= \frac{2}{\sqrt{\al}}\int^t_\infty \sqrt{-\dot H(\tau)}\,d
\tau.
\ee
Remind that we assume $\si(t)\to 0$ at $t\to \infty$.
Equations~(\ref{meq}) and ({\ref{Hsi}) supersede the
pair of the original evolution equations~(\ref{eveq}).

\paragraph{Self-consistent solution}

Let us put in what follows $\Lambda_{\rm s}\equiv 0$. This will be
justified afterwards. Iterating
eq.~(\ref{meq}), with $\Lambda$ considered as a perturbation, we
can get the solution with any desired accuracy. In particular,
substituting into the r.h.s.\ of eq.~(\ref{meq}) the solution
$H=\al/t$ from the zeroth approximation ($\Lambda=0$) we get  the
first approximation as follows
\be\label{H'}
H^2=\Big(\frac{\al}{t}\Big)^2+\frac{\Lambda}{3}+
\cases{{\cal O}(1)& {\rm at} $t\to 0$ , \cr
{\cal O}(1/t^3) & {\rm at}  t $\to \infty $,}
\ee
or otherwise
\be\label{H}
H^2=\frac{\al^2}{t_\Lambda^2}\Big((t_\Lambda/t)^2+1\Big)\simeq
\cases{\al^2/t^2& {\rm at} $t/t_\Lambda<1$,\cr
\al^2/t_\Lambda^2& {\rm at} $t/t_\Lambda>1$,}
\ee
with
\be
t_\Lambda=\frac{\al}{\sqrt{\Lambda/3}}
\ee
being the characteristic time of the evolution of the Universe.
Numerically, $t_\Lambda\sim 10^{10}$yr is of order the age of the
Universe. Equation~(\ref{H}) is the basis for the qualitative
discussion in what follows. 

Integrating  eq.~(\ref{H})  we get the scale factor as follows:
\be\label{at}
\ln \frac{a}{a_0}=
\al\bigg[\frac{t}{t_\Lambda}\sqrt{\Big(\frac{t_\Lambda}{t}\Big)^2+1}
-\ln\bigg(\frac{t_\Lambda}{t}+\sqrt{
\Big(\frac{t_\Lambda}{t}\Big)^2+1}\, \bigg)\bigg]
\sim\cases{\al \ln (t/t_\Lambda)
&{\rm at} $t/t_\Lambda< 1$, \cr
\al t/t_\Lambda  &{\rm at} $t/t_\Lambda> 1$,}
\ee
where $a_0$ is an integration constant.\footnote{To
phenomenologically account  for
the effect of the initial inflation  we could formally shift the
origin of time: $t\to t+t_0$, with $t_0>0$.}
Explicitly, the scale factor looks like
\be\label{asim}
a\sim\cases{(t/t_\Lambda)^\al
&{\rm at} $t/t_\Lambda< 1$, \cr
\exp(\al t/t_\Lambda)   &{\rm at} $t/t_\Lambda> 1$,
}
\ee
with $t_\Lambda$ bordering thus the epoch of the the power law
expansion from the epoch of the exponential expansion.
Equation~(\ref{at}) gives the two-parametric
representation for the scale factor of the acceleratedly expanding
Universe after the initial period.   

With account for eq.~(\ref{intsi}) the $\si$-field behaves as
\be\label{intsi'}
\si=-\int_{(t/t_\Lambda)^2}^\infty\frac{d\xi}{\xi(1+\xi)^{1/4}}
\sim
\cases{2\ln(t/t_\Lambda) &{\rm at}
$t/t_\Lambda<1$,\cr
-4\sqrt{t_\Lambda/t}& {\rm at} $t/t_\Lambda>1$.}
\ee
Note that at $t_\Lambda\to \infty$ or, equivalently,
$\Lambda\to 0$ the integral above diverges and 
the  $\si$-field can not be normalized properly.
$\Lambda \neq 0$ is thus necessary as a regulator in the theory.
Now, the consistency condition looks like
\be\label{tildea}
\tilde a=a\exp{(-\si/3)} \sim \cases{
(t/t_\Lambda)^{\al-2/3} &{\rm at} $t/t_\Lambda<1$,\cr
\exp{(\al t/t_\Lambda})& {\rm at} $t/t_\Lambda>1$.}
\ee
Clearly, $\Lambda$ should be already  presupposed in 
$\tilde a$. Note that in the case $\al=2/3$, the parameter $\tilde a$
is approximately constant at $t/t_\Lambda<1$.

Substituting equations~(\ref{H}) and (\ref{Hsi}) into the first line
of eq.~(\ref{rhop}) we can explicitly verify that
\be\label{rhos}
\frac{1}{m^2_{\rm P}}\rho_{\rm s}=\frac{3\al^2}{t^2}.
\ee
This is to be anticipated already from the relation
$\rho_\Lambda/m^2_{\rm P}=\Lambda$, as well as eq.~(\ref{H'}) and
the first line of eq.~(\ref{eveq}). Clearly, $\rho_{\rm s}$ is
positive.
At $t/t_\Lambda<1$ we get from equations~(\ref{asim}) and  
(\ref{rhos}) that $\rho_{\rm s}a^3\sim t^{3\al-2}$.  
In the case  $\al=2/3$, we have $\rho_{\rm s}\sim 1/a^{3}$  as it
should be for the true CDM.
On the other hand, the pressure of the scalar gravitons is as follows
\be\label{p}
\frac{1}{m_{\rm P}^2} p_{\rm s}
=-\frac{3\al^2}{t^2}+\frac{2\al}{t_\Lambda^2}\frac{(t_\Lambda/t)^3}
{\sqrt{(t_\Lambda/t)^2+1}}
\simeq\cases{3\al(2/3-\al)/t^2- \al/t_\Lambda^2&{\rm at} $t/t_\Lambda<
1$,\cr 
(2\al/t_\Lambda^2)(t_\Lambda/t)^3 &{\rm at} $t/t_\Lambda> 1$.}
\ee
At the same conditions as before,  the pressure is 
$p_{\rm s}/{m_{\rm P}^2} =-\Lambda $/2, being near constant though
not zero as it  should be anticipated for the true CDM. Nevertheless,
we see that the value $\al=2/3$ is exceptional
in many respects. Conceivably, such a value is distinguished by a more
fundamental theory.

Introducing the critical energy density $\rho_{\rm c}=
3m_{\rm P}^2 H^2$, we get for the partial energy densities 
$\Omega_{\rm  s}=\rho_{\rm s}/\rho_{\rm c}$ 
and $\Omega_\Lambda=\rho_\Lambda/ \rho_{\rm c} $, respectively, 
of the scalar gravitons and the $\Lambda$-term the following:
\be
\Omega_{\rm  s} =\frac{1}{1+(t/t_\Lambda)^2}\simeq
\cases{1-(t/t_\Lambda)^2
&{\rm at} $t/t_\Lambda< 1$, \cr  
(t_\Lambda/t)^2&{\rm at} $t/t_\Lambda> 1$, }
\ee
with $\Omega_\Lambda= 1-\Omega_{\rm  s}$. Note that $\Omega_{\rm  s}=
\Omega_\Lambda =  1/2$ at $t/t_\Lambda=1$.  Presently, we have
$\Omega_{\rm  s}/\Omega_\Lambda\simeq 1/3$ and
thus the respective time $t$, in the neglect by the effect of the
initial inflation, is somewhat larger~$t_\Lambda$.  

Finally, the condition $\Lambda_{\rm s}=0$ adopted earlier  can be
justified as follows.
First of all, $\Lambda_{\rm s}$ is indeed negligible at $t\to \infty$ 
due to $\si\to 0$ and $\Lambda_{\rm s}(0)=0$. On the other
hand, at $t\sim t_\Lambda$ we have $|\si|\sim 1$ and hence
$\Lambda_{\rm s}\sim m_{\rm s}^2$. For $\Lambda_{\rm s}$
to be negligible in this region, too, we should require
$m_{\rm s}\leq\sqrt{\Lambda}\sim 1/t_\Lambda$. 
Nevertheless, in the early period of evolution when $|\si|>1$ the
contribution of $\Lambda_{\rm s}$ may
be significant. The parameter $\al= 2/3$ being fixed the theory
may be terminated just by two mass parameters: the ultraviolet 
$m_{\rm P}$ and  the infrared $t_\Lambda^{-1} \sim\sqrt \Lambda$ or, 
otherwise,~$m_{\rm s}$.

\section{Conclusion}

To conclude, let us recapitulate the proposed solution to the DM and
DE problems in the context of the evolution of the Universe. According
to the viewpoint adopted, there is neither true DM nor DE in the
Universe (at least, in a sizable amount). Instead,
the field $\si$ of the scalar
graviton serves as a common source of both the DM and DE of
the gravitational origin. DM is represented by the derivative
contribution of $\si$, with DE being reflected by the
derivativeless contribution. In this, the constant part of
the latter contribution corresponds to the conventional
$\Lambda$-term, while the $\si$-dependent part corresponds to DE.
The latter is less important than  the $\Lambda$-term  at
present, becoming conceivably more crucial at the early time. 

The self-consistent evolution of the Universe  may be considered as
the transition of the ``sea'' of the scalar gravitons,
produced in the early period, from the excited state with
$\vert\si\vert> 1$ to the ground state with $\si=0$. The ground state
is characterized by the cosmological constant $\Lambda$ which, in
turn, predetermines the characteristic evolution time of the Universe,
$t_\Lambda\sim 1/\sqrt\Lambda$. The scenario is in the possession to
naturally describe the accelerated expansion of the spatially flat
Universe, correctly emulating thus the
conventional LCDM model. The more complete study of the scenario, the
initial period of the evolution including, is in order.

The author is grateful to O.~V.~Zenin for the useful discussions.

\end{document}